\documentclass[journal,10pt]{IEEEtran}
\usepackage{amsmath,amsfonts,amssymb,graphicx}
\usepackage{placeins}
\usepackage{multirow,makecell}
\usepackage{MnSymbol}
\usepackage{bm}

\usepackage{flushend}
\usepackage{graphicx,amsthm}
\usepackage{enumerate}
\usepackage{comment,amssymb,epsfig,amsfonts,verbatim}
\usepackage{psfrag}
\usepackage{graphicx,times}
\usepackage{cite}
\usepackage{units}
\usepackage{xcolor}
\usepackage{eucal}
\usepackage{stfloats}
\usepackage{subfig}
\usepackage{calrsfs}

 \newcommand{\cblue}{\textcolor{black!90!black}}
 \newcommand{\alphab}{\boldsymbol{\alpha}}
\newcommand{\mub}{\boldsymbol{\mu}}

\usepackage[ruled,vlined]{algorithm2e}

\theoremstyle{definition}

\def\n{{\mathbf n}}

\def\mcN{{\CMcal N}}

\def\x{{\bf x}}
\def\s{{\bf s}}

\def\z{{\bf z}}

\def\mcR{\CMcal{R}}
\def\mcS{\CMcal{S}}

\newcommand{\Var}{\mathrm{Var}}

\newcommand{\Sigmab}{{\bm \Sigma}}

\newcommand{\omegab}{{\bm \omega}}
\newcommand{\gammab}{{\bm \gamma}}

\title{Multiple Importance Sampling for Efficient \\Symbol Error Rate Estimation}

\author{V\'ictor~Elvira,~\IEEEmembership{Member,~IEEE,} and~Ignacio~Santamar\'ia,~\IEEEmembership{Senior Member,~IEEE,}
\thanks{V.~Elvira is with IMT Lille Douai \& CRISTAL (UMR 9189), Lille, France (e-mail: victor.elvira@imt-lille-douai.fr).}
\thanks{I.~Santamaria is with the Department of Communications Engineering, University of Cantabria, Santander, Spain (e-mail: i.santamaria@unican.es).}
\thanks{The work of V. Elvira was partially supported by \emph{Agence Nationale de la Recherche} of France under PISCES project (ANR-17-CE40-0031-01) and the French-American Fulbright Commission. The work of I. Santamaria was partly supported by the Ministerio de Econom\'{i}a y Competitividad (MINECO) of Spain, and AEI/FEDER funds of the E.U., under grant TEC2016-75067-C4-4-R (CARMEN).}}

\begin{document}
%
\maketitle
\begin{abstract}

Digital constellations formed by hexagonal or other non-square two-dimensional lattices are often used in advanced digital communication systems. The integrals required to evaluate the symbol error rate (SER) of these constellations in the presence of Gaussian noise are in general difficult to compute in closed-form, and therefore Monte Carlo simulation is typically used to estimate the SER. However, naive Monte Carlo simulation can be very inefficient and requires very long simulation runs, especially at high signal-to-noise ratios. In this letter, we adapt a recently proposed multiple importance sampling (MIS) technique, called \cblue{ALOE} (for ``At Least One rare Event''), to this problem. Conditioned to a transmitted symbol, an error (or rare event) occurs when the observation falls in a union of half-spaces or, equivalently, outside a given polytope. The proposal distribution for \cblue{ALOE} samples the system conditionally on an error taking place, which makes it more efficient than other importance sampling techniques. \cblue{ALOE} provides unbiased SER estimates with simulation times orders of magnitude shorter than conventional Monte Carlo.

\end{abstract}

\section{Introduction}
\label{sec:intro}
 
In several communication systems the transmitted signal belongs to a non-square 2D lattice. For instance, hexagonal constellations are the densest packing of regularly spaced points in 2D and, as the number of constellation points grows to infinity, they also minimize the probability of error in Gaussian noise under an average power constraint being therefore optimal \cite{Forney98}. Other examples of constellations that provide non-rectangular decision boundaries are the $\theta$ quadrature amplitude modulation (QAM) family \cite{Pappi10}, and the recently proposed family of improper constellations in \cite{Santamaria2018}.

When these constellations are used, maximum likelihood decoding amounts to finding the closest lattice point to a given noisy observation, $x$, which is known to be NP-hard for generic channels \cite{Grotschelbook, Caire03} \cblue{meaning that the problem complexity is exponential on the dimension of the lattice \cite{Zamirbook}}. Further, to evaluate the symbol error rate probability conditioned to a given constellation point we need to compute the probability that the observation $x$ lies outside a polytope defining the corresponding Voronoi region. The resulting integrals are in general difficult to compute in closed form for arbitrary decision regions, and typically one resorts to standard Monte Carlo simulations for performance evaluation, which can be very time consuming especially when the number of symbols of the constellation is large and the targeted symbol error rate (SER) is below $10^{-6}$.

As an alternative to naive Monte Carlo, importance sampling (IS) has been used as a method for variance reduction in SER or BER simulations in a wide range of scenarios since the late seventies \cite{Balaban76, Shanmugam80,Lu88,Townsend97} (for a review we refer the reader to \cite{Smith97}). Despite this, many digital communication researchers are still unaware of the potential benefits of IS techniques to characterize the statistical performance of digital communication systems. 
 
The basic IS methodology samples from a proposal distribution that increases the number of errors during simulation, and then weights the samples by the ratios of the target to the proposal densities \cite{Bugallo17}.
However, designing a good proposal with high density in regions where samples should be drawn might not be easy, and typically is problem dependent. For instance, a biased channel distribution is proposed in \cite{Jacobs14} for the simulation of orthogonal space-time block codes (OSTBCs) on Nakagami channels. Other recent work is \cite{Alouini18}, where IS is used for the estimation of the outage probability of multi-antenna receivers with generalized selection combining. 
A general approach is using multiple importance sampling (MIS), where several proposals are used for simulating the samples \cite{veach1995optimally,owen2000safe,elvira15,elvira2016heretical}.

In this letter, we apply a recently proposed multiple importance sampling technique called \cblue{ALOE} (``At Least One rare Event'') \cite{owen2017importance} to estimate the SER in additive white Gaussian noise (AWGN) channels. 

The conditional symbol error probability is the integral of a Gaussian outside a polytope formed by the intersection of $K$ hyperplanes; or, in other words, the integral over the union of the half-spaces formed by these hyperplanes.
\cblue{ALOE} estimates this integral by taking samples from a mixture of $K$ truncated Gaussians where errors take place. The method can be easily extended to arbitrary 2D constellations and provides unbiased SER estimated with bounded variance.

\section{Efficient SER estimation}
\label{sec:IS}

\subsection{Problem statement}
\begin{figure}
	\centering
	\subfloat[]{	
		\includegraphics[width=\columnwidth]{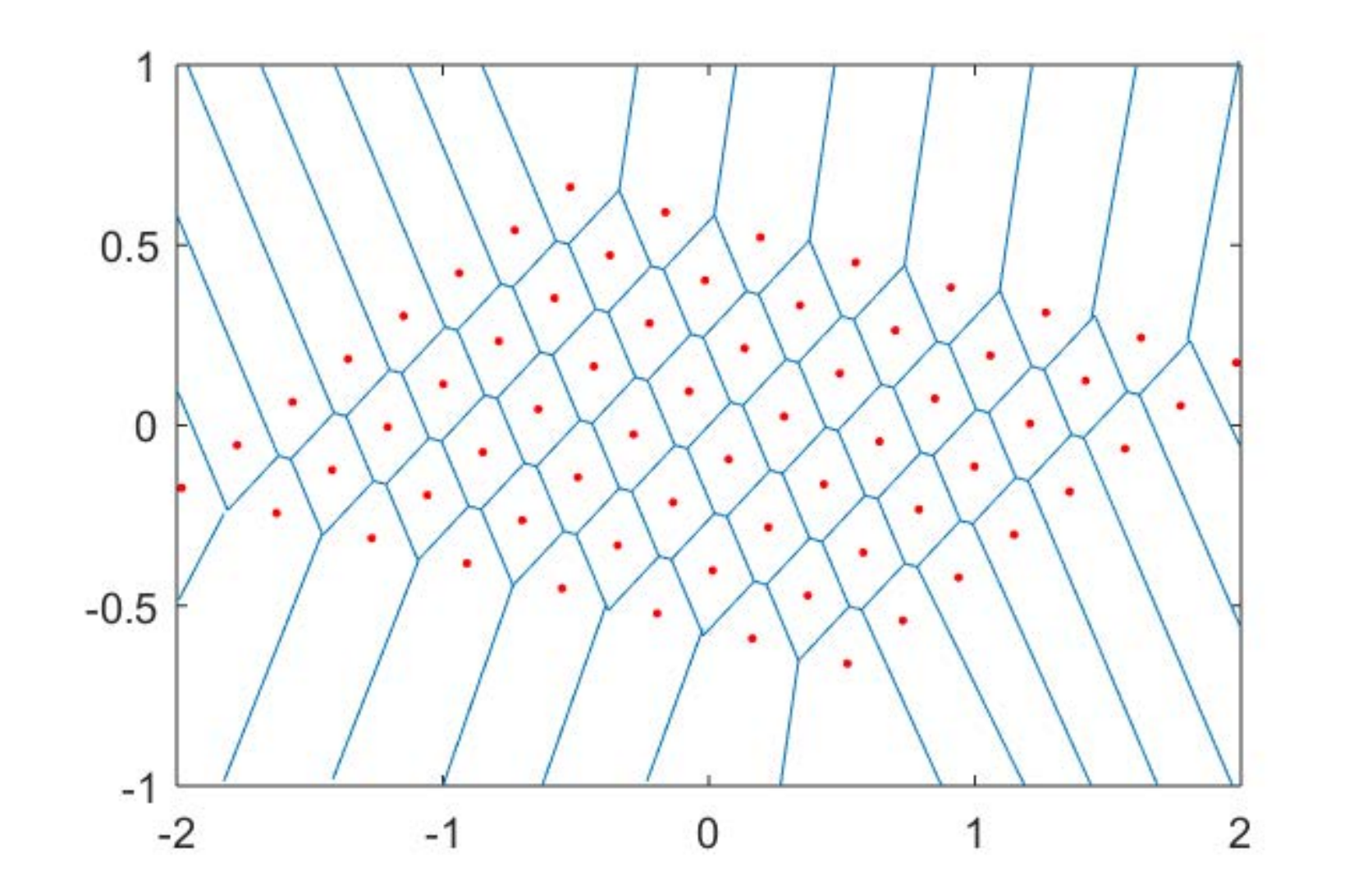}
		\label{Fig1a}}	
	\vspace{1em} 
		\subfloat[]{
	
		\includegraphics[width=\columnwidth]{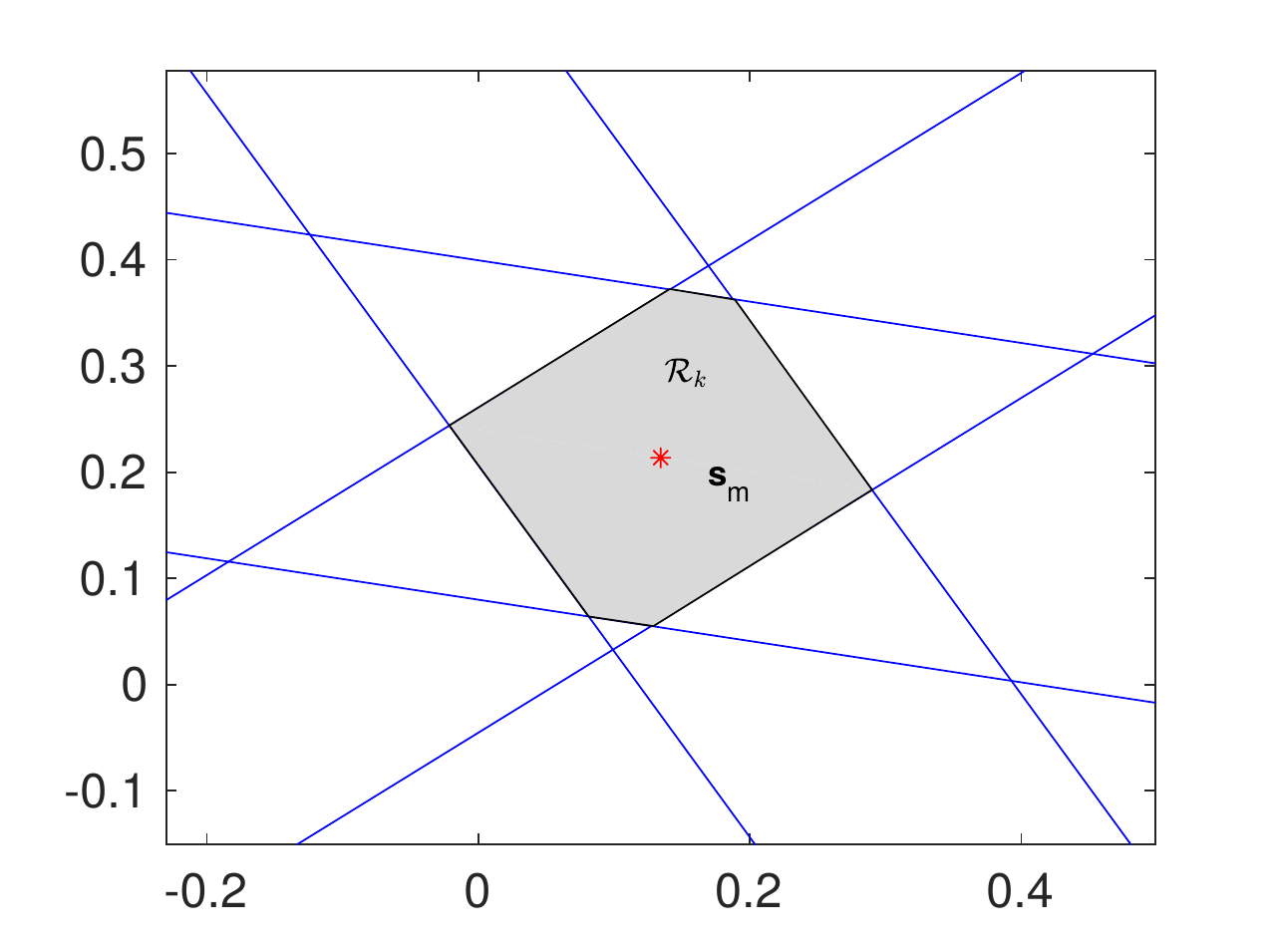} \label{Fig1b}}	
	\caption{a) 2D lattice for an improper constellation with $M=64$ and $\kappa = 0.8$. The decision regions are also depicted, and b) Zoom of the decision region for a given transmitted symbol $\s_m = 0.1343 + 0.2136j$.} 
	
\end{figure}

Our motivating problem is the efficient SER estimation of digital constellations like the one depicted in Fig. \ref{Fig1a}. Assume that symbol $\s_m$, marked with a star in Fig. \ref{Fig1b}, is transmitted.
The observation $\x = \s_m + \n$ is the symbol perturbed by additive Gaussian noise. The conditional probability of error is the integral of a Gaussian distribution centered at $\s_m$ outside the shaded decision region
in Fig. \ref{Fig1b}: $p_m \triangleq P(e|\s_m) = \mathbb{P}(\x \notin {\mcR}_m|\s_m)$, and, assuming that all symbols are transmitted with equal probability, the symbol error probability is
\begin{equation}
P_e = \frac{1}{M} \sum_{m=1}^M p_m. 
\label{eq:SER}
\end{equation}

Our goal is to develop an efficient multiple important sampling (MIS) scheme to estimate $p_m$, $m=1,\ldots,M$ in (\ref{eq:SER}).   

For notational simplicity let us denote the transmitted symbols as $\s$, and let $\mcR$ be its associated Voronoi region, which is a polytope defined by the intersection of finitely many hyperplanes in $\mathbb{R}^2$. The probability of interest can be then expressed as
\begin{equation}
p = \int_{\mathbb{R}^2} \mathbb{I}_{\overline{\mcR}}(\x) \tilde \pi(\x) d\x
\label{eq_integral}
\end{equation}
where $\tilde \pi(\x) \triangleq \mcN(\s,\sigma^2 {\bf I})$ is the Gaussian distribution of the observation, and $\mathbb{I}_{\overline{\mcR}}(\x)$ is the indicator function taking value 1 for all $\x\nin \mcR$, i.e., out of the shaded decision region
in Fig. \ref{Fig1b}. The integral in Eq. \eqref{eq_integral} is intractable in the general case where $\mcR$ is an arbitrary polytope. However, for some constellations such as QAM modulations, the integral can of course be obtained in a closed-form, and there is no need to resort to any simulation procedure. 

A naive Monte Carlo estimator of $p$ simulates $N$ samples from $ \tilde \pi(\x)$ and approximates \eqref{eq_integral} as
\begin{equation}
\widehat{p}^{\, \, \, \text{(MC)}} = \frac{1}{N} \sum_{n=1}^N   \mathbb{I}_{\overline{\mcR}}(\x_n),
\label{eq_naive_MC}
\end{equation}
i.e., it simulates from the observation distribution and counts the rate of observations that are out of $\mcR$. This estimator is unbiased, but its efficiency decays dramatically when $p$ is small.

\subsection{Importance Sampling}

Importance sampling (IS) is a more advanced Monte Carlo methodology, used when sampling from $ \tilde \pi(\x)$ is either not possible or not efficient. The $N$ samples are simulated instead from a so-called \emph{proposal} distribution, $q(\x)$, and the estimator of $p$ is built as
 \begin{equation}
\widehat{p}^{\, \, \, \text{(IS)}}= \frac{1}{N} \sum_{n=1}^N w_n \mathbb{I}_{\overline{\mcR}}(\x_n), \qquad \x_n \sim q(x),\qquad n=1,...,N,\label{eq_UIS}
\end{equation}
where $w_n = \frac{\tilde \pi(\x_n)}{q(\x_n)}$ is the importance weight. \cblue{Note that} $\widehat{p}^{\, \, \, \text{(IS)}}$ is an unbiased and consistent estimator of $p$.

The variance of the estimator $\widehat{p}^{\, \, \, \text{(IS)}}$ in Eq. \eqref{eq_UIS} is given by
\begin{equation}
\Var_{q}(\widehat{p}^{\, \, \, \text{(IS)}}) = \frac{1}{N} \int \frac{\mathbb{I}_{\overline{\mcR}}(\x)  \tilde \pi^2(\x) }{q(\x)}d\x - \frac{p^2}{N}.
\label{eq_var_is}
\end{equation}
It can be shown that the optimal proposal, $q^*$, that minimizes the variance is $q^*(\x) \propto \mathbb{I}_{\overline{\mcR}}(\x) \cdot  \tilde \pi(\x)$. Intuitively, the performance is poor when the targeted integrand $\mathbb{I}_{\overline{\mcR}}(\x)\cdot \tilde\pi(\x)$ and the distribution that generates the samples have a large mismatch. In our problem, the performance is very bad when we find few or no errors for the simulated observation. This explains the very poor performance of the naive Monte Carlo $\widehat{p}^{\, \, \, \text{(MC)}}$, whose variance can be obtained as a particular case of \eqref{eq_var_is} as

\begin{align}
\Var_{\tilde \pi}(\widehat{p}^{\, \, \, \text{(MC)}}) &= \frac{1}{N}\left(p - p^2 \right),
\end{align}
with a relative root mean square error (RRMSE) of

\begin{align}
\text{RRMSE}(\widehat{p}^{\, \, \, \text{(MC)}}) &= \frac{ \sqrt{\Var(\widehat{p}^{\, \, \, \text{(MC)}})} }{p} \\ &= \frac{1}{\sqrt{N} } \frac{\sqrt{p - p^2 } }{p} \\ &= \frac{1}{\sqrt{N} } \sqrt{\frac{1}{p} - 1}.
\label{eq:relerror}
\end{align}
Note that $N$ needs to be inversely proportional to the true value $p$ in order to provide a reliable estimate, which explains why $N$ must be huge for high signal-to-noise ratio (SNR).


\subsection{Multiple Importance Sampling}
\label{sec:mis}
In the case of IS, it is difficult to find a good unique proposal with a low mismatch w.r.t. the multimodal product $ \mathbb{I}_{\overline{\mcR}}(\x) \cdot  \tilde \pi(\x)$. 
Multiple importance sampling (MIS) is a natural extension of IS in this setup, allowing for the simulation from a set of $K$ proposals, $\{q_k(\x) \}_{k=1}^{K}$, instead of just one \cite{elvira15}. 
However, the extension from one to several proposals is not straightforward and many sampling and weighting schemes can be devised (see \cite{elvira2018generalized} for a review). A conventional way to proceed \cblue{is to simulate} from the mixture proposal as
\begin{equation} 
\x_{n} \sim q_{{\alphab}} = \sum_{k=1}^K \alpha_{k} q_{k}(\x), \qquad n=1,...,N,
\label{eq_mixture_proposal}
\end{equation}
where $\alphab = [\alpha_1,...,\alpha_K]$ is a simplex vector with all non-negative weights in the mixture such that $\sum_{k=1}^K \alpha_k= 1$. The MIS extension of Eq. \eqref{eq_UIS} is
\begin{equation} 
\widehat{p}^{\, \, \, \text{(MIS)}}= \frac{1}{N}\sum_{n=1}^N \frac{ \mathbb{I}_{\overline{\mcR}}(\x_n)\tilde \pi(\x_n)}{q_{\alphab}(\x_n)}.
\label{eq_est_mis_generic}
\end{equation}

We recall that we aim at integrating $\tilde \pi(\x)$ in region $\overline{\mcR}$. This region can be described as the union of all half-spaces in $\mathbb{R}^2$ generated by the $K$ hyperplanes that define the border of $\mcR$ (e.g, $K=6$ for the symbol in Fig. \ref{Fig1b}). 
More precisely, $\overline{\mcR} = \bigcup\limits_{k=1}^{K} \mcS_{k}$, where $\mcS_k = \{\x \in \mathbb{R}^2 \, |  \, \x^T \gammab_k \geq \beta_k\}$ is the half-space defined by the $k$-th hyperplane, which is parametrized by $\gammab_k$ and $\beta_k$.

\begin{figure}[htb]
    \begin{center}
        \includegraphics[width=0.95\columnwidth]{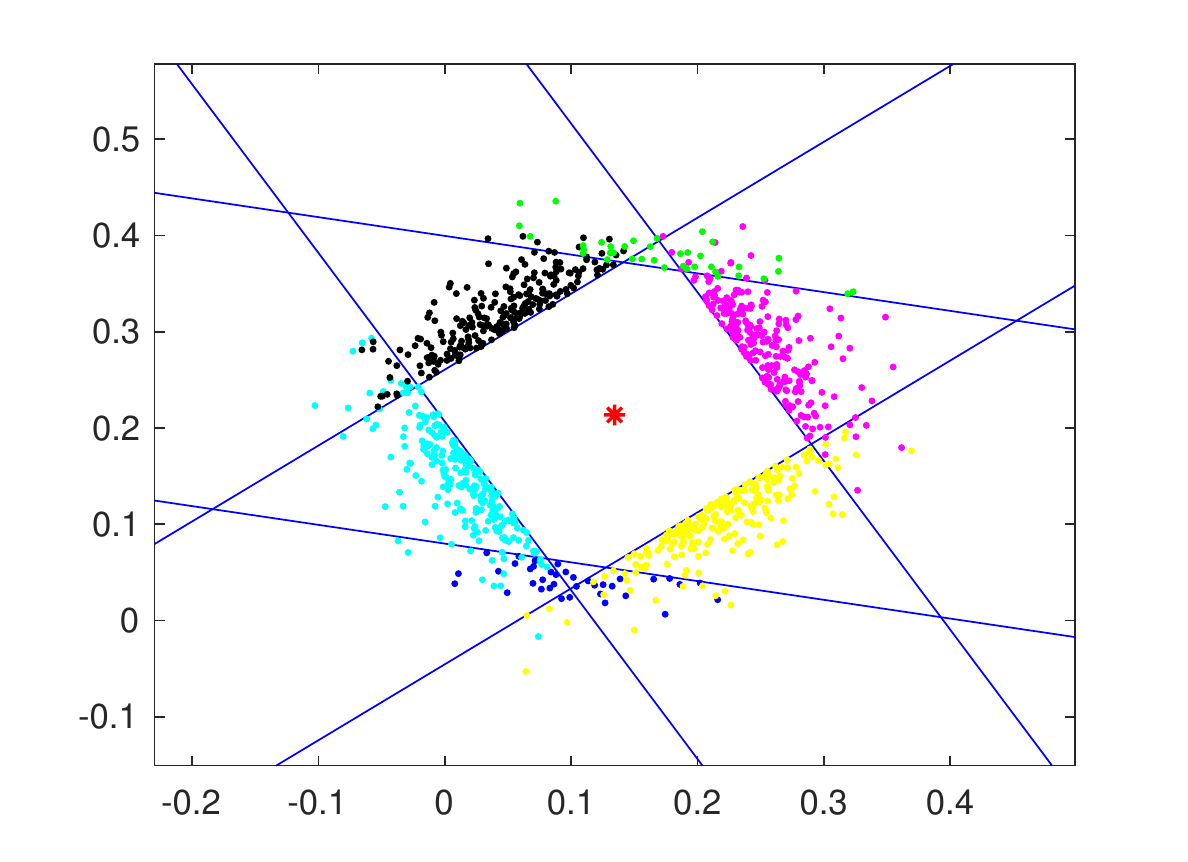}
    \end{center}
    \caption{The red asterisk represents the symbol $\s_m = 0.1343 + 0.2136j$. The blue solid lines represent the $K=6$ hyperplanes that define the half-spaces, whose union describes $\overline{\mcR}$. The points represent the samples simulated from the mixture proposal $q_{{\alphab}}$ (each component in a different color).}
    \label{fig:MIS_plot}
\end{figure}

We follow the choice of proposals in \cite{owen2017importance}. First, the number of proposals, $K$, is equal to the number of hyperplanes, being each proposal a truncated version of the target distribution (a Gaussian centered at the received symbol) beyond each hyperplane, i.e., $q_k(\x) = \frac{\mathbb{I}_{\mcS_k(\x)}\tilde\pi(\x)}{P_k}$, where $P_k = \int \mathbb{I}_{\mcS_k(\x)}\tilde\pi(\x)d\x$ is the integral of the target distribution beyond the hyperplane (the procedure for the efficient simulation from a generic truncated Gaussian distribution is described in Appendices \ref{ap1} and \ref{ap2}). For reasons that will become apparent shortly, it is useful to define $\overline{p} = \sum_{k=1}^K P_k$, which is an upper union bound of $p$.\footnote{Note that the bound becomes an equality when $S_i \cap S_j = \emptyset$, for all $i\neq j$.} Then, Eq. \eqref{eq_est_mis_generic} yields
\begin{align} 
\widehat{p}^{\, \, \, \text{(MIS)}}&= \frac{1}{N}\sum_{n=1}^N \frac{ \mathbb{I}_{\overline{\mcR}}(\x_n)\tilde \pi(\x_n)}{q_{\alphab}(\x_n)} \\
&=\frac{1}{N}\sum_{n=1}^N \frac{\mathbb{I}_{\overline{\mcR}}(\x_n)}{\sum_{k=1}^K \alpha_k \mathbb{I}_{\mcS_k}(\x_n) P_k^{-1} }.
\label{eq_est_mis_1}
\end{align}
The weight of each proposal in the mixture defined in Eq. \eqref{eq_mixture_proposal} is chosen as $\alpha_k = P_k/\bar{p}$, for $k=1,...,K$. Then, Eq. \eqref{eq_est_mis_1} yields
\begin{align}
&=\frac{\overline{p}}{N}\sum_{n=1}^N \frac{\mathbb{I}_{\overline{\mcR}}(\x_n)}{\sum_{k=1}^K  \mathbb{I}_{\mcS_k}(\x_n) } \\
&= \frac{\overline{p}}{N}\sum_{n=1}^N \frac{1}{C(\x_n) },
\label{eq_est_mis_2}
\end{align}
where $C(\x_n)=\sum_{k=1}^K  \mathbb{I}_{\mcS_k}(\x_n) $ is the number of half-spaces $\mcS_k$ where $\x_n$ is present. Note that we have used $\mathbb{I}_{\overline{\mcR}}(\x_n)=1$, since all samples are generated in $\overline{\mcR}$. Fig. \ref{fig:MIS_plot} shows samples generated from the proposal corresponding to the symbol $\s_m = 0.1343 + 0.2136j$ in the constellation of Fig. \ref{Fig1a}. Note that there are $K=6$ proposals in the mixture, and the samples of each proposal are represented by a different color.

\subsection{Theoretical guarantees of the MIS estimator}

The variance of the estimator \eqref{eq_est_mis_2} can be bounded by \cite{owen2017importance}
\begin{align}
\Var(\widehat{p}^{\, \, \, \text{(MIS)}})&\leq \frac{p(\overline{p}-p)}{N}.
\label{eq:varMIS}
\end{align}
First, note that when the upper union bound $\bar{p}$ gets closer to the true value $p$, the variance of the MIS estimator goes to zero. This is of special interest in our application since the performance of the raw Monte Carlo estimate deteriorates for high SNRs (i.e., when $p$ is small). Conversely, the MIS estimator gets better for high SNR since the upper bound $\bar{p}$ becomes tighter. The reason is that most errors fall very close to the separating hyperplane, and hence in just only one region $S_k$ (i.e., $C(\x_n)=1$ for all $\x_n$). Note that in the limit, when all $C(\x_n)=1$, the estimator \eqref{eq_est_mis_2} is always $\overline{p}$.

Let us compare the variance bound of the MIS estimator and the closed-form variance of raw Monte Carlo. We look for values of $\overline{p}$ where
\begin{align}
\frac{p(\overline{p}-p)}{N} &\leq \frac{1}{N}\left(p - p^2 \right),
\end{align}
which turns to be $\overline{p} \leq 1$. Note that for moderate SNRs, the upper bound $\overline{p}$ is smaller than 1. Only for very low SNRs, $\overline{p} \geq 1$, which is not a problem since (a) we are comparing the MIS upper bound (not the variance itself), and (b) the low SNR scenario is not challenging and both MIS and raw Monte Carlo estimators are very accurate with a few samples. 

Eq. (\ref{eq:varMIS}) gives us an upper bound for the variance of the conditional probability of error of an arbitrary symbol of the constellation that we denoted for simplicity as ${\bf s}$. To estimate the symbol error probability in Eq. (\ref{eq:SER}) we run $M$ independent \cblue{ALOEs}, then, the variance of the $P_e$ estimator can be bounded as
\begin{align}
\Var(\widehat{P}_e^{\, \, \, \text{(MIS)}})&\leq \frac{1}{M^2 N} \sum_{m=1}^Mp_m(\overline{p}_m-p_m).
\label{eq:varMISPe}
\end{align} 
Note that several symbols in Fig. \ref{Fig1a} have identical Voronoi regions, which can be exploited to avoid approximating all corresponding error probabilities. Moreover, one can allocate different number of samples, $N_m$, for the approximation of the error probability associated to each of the $M$ symbols. However, this fine tuning goes beyond the scope of this paper.

\section{Simulation Results}
\label{sec:simulations}

In this section we estimate the SER of arbitrary 2D digital constellations with the proposed MIS technique. As an example, we consider the family of improper constellations proposed in \cite{Santamaria2018}, whose two-dimensional signal points belong to a non-square lattice. Improper signals, which are correlated with their complex conjugate, have proven beneficial in several interference-limited wireless networks such as the interference channel\cite{Cadambe2010,Ho2012,LagenMorancho2016}, Z-interference channel \cite{Lameiro2017}, underlay cognitive radio networks \cite{Lameiro15, Amin16}, and relay channels \cite{Gaafar2016letter}.

In all these scenarios, we wish to transmit a digital communications signal with a given circularity coefficient, which is the parameter that determines the degree of impropriety of the signal. The circularity coefficient of a zero-mean complex random variable $X$, is defined as \cite{Schreierbook}
\begin{equation}
	\kappa=\frac{\left|\operatorname{E}[X^2]\right|}{\operatorname{E}[|X|^2]}, \quad \quad 0\leq\kappa\leq1 \label{eq:kdef}
\end{equation}

\begin{figure}[htb]
    \begin{center}
        \includegraphics[width=0.989\columnwidth]{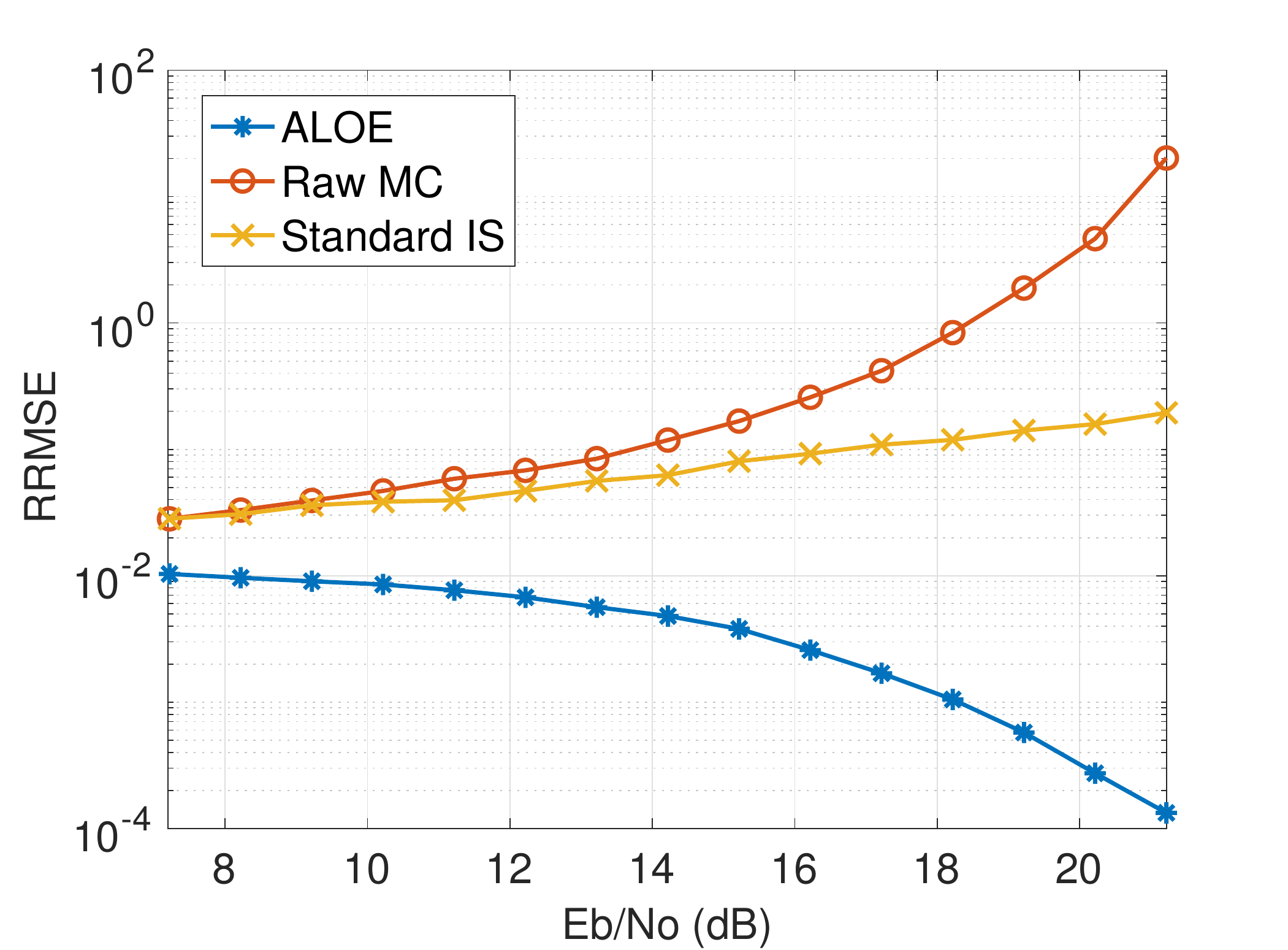}
    \end{center}
    \caption{Relative error variance of SER curves estimated with ALOE, Raw-MC, and a standard IS algorithm.}
    \label{fig:REcurve}
\end{figure}

We consider a constellation with $M=64$ symbols and circularity coefficient $\kappa = 0.8$, whose signal points are shown in Fig. \ref{Fig1a}. We estimate the SER curve using \cblue{ALOE} and conventional Monte Carlo transmitting only $N = 1280$ symbols at each $E_b/N_0$ value, \cblue{i.e., each estimator uses $N = 1280$ samples (we transmit $20$ times each of the $M=64$ symbols).} 
\cblue{Moreover, we also compare with a standard IS algorithm with a unique Gaussian proposal centered at each symbol $\s$, and variance $\alpha^2 \sigma^2 {\bf I}$, with $\alpha>1$, i.e., the proposal is overdispersed w.r.t. the target $\tilde \pi(\x)$, since $\mathbb{I}_{\overline{\mcR}}(\x) \tilde \pi(\x)$ is clearly more dispersed than $\tilde \pi(\x)$ (see the discussion about the optimal proposal in \cite{Robert04,mcbook} and the use of overdispersed IS proposals in \cite{ruiz2016overdispersed}). We try a grid of values of $\alpha\in\{1,1.5,2,...,5 \}$, and we select the case with the smallest RRMSE.} 
The experiment is repeated 200 times and the \cblue{RRMSEs} defined in (\ref{eq:relerror}) are depicted in Fig. \ref{fig:REcurve}. For high $E_b/N_0$ values, \cblue{ALOE} provides four to five orders of magnitude improvement over the usual naive Monte Carlo.

\section{Conclusions}

A multiple importance sampling technique called \cblue{ALOE} is used in this paper to estimate the symbol error rate probability of 2D constellations designed over arbitrary lattices. \cblue{ALOE} is extremely efficient to estimate the integral of a Gaussian in a region defined by a union of half-spaces, which is precisely the error event in a digital communications system. At high signal-to-noise ratios, \cblue{ALOE} provides orders of magnitude speedup with respect to conventional Monte Carlo simulation. As future work, we plan to extend the proposed SER estimation technique to other noise distributions, {\cblue {to higher dimensional lattices originated by multi-antenna systems}}, as well as to fading channels. 
constellations with enhanced features.

\section*{Acknowledgments}
The authors want thank Art B. Owen for his insightful comments.

\appendix
\subsection{Simulation from a truncated Gaussian $\mcN(\textbf{0},\textbf{I})$}\label{ap1}
The proposals in the MIS implementation are truncated Gaussian distributions. Let us first describe the simulation of a truncated Gaussian $\mcN(\textbf{0},\textbf{I})$ in the half-space described by $\x^T\omegab\geq \tau$, which first proceeds by simulating the sample $\x$ from the complementary half-space (i.e., $\x^T\omegab < \tau$), and then delivering the $-\x$ for numerical stability. The algorithm described in \cite{owen2017importance} proceeds as follows:
\begin{enumerate}
\item Simulate  $\z \sim \mcN (\textbf{0},\textbf{I})$
\item Simulate $u \sim {\CMcal U } ( 0,1 )$ 
\item Let $y = \Phi ^ { - 1 } ( u \Phi ( - \tau ) )$, where $\Phi(\tau)$ denotes the cumulative distribution function for the standard Gaussian
\item Let $\mathbf { x } = \omegab y + \left( \textbf{I} - \omegab \omegab ^ { T } \right) \z$ 
\item Output $\x = - \x$ 
\end{enumerate}
\subsection{Extension to a generic truncated Gaussian $\mcN(\mub,\Sigmab)$}\label{ap2}
When the target distribution is a generic truncated Gaussian $\mcN(\mub,\Sigmab)$ that must be integrated over the union of the half-spaces described by the hyperplanes $\{ \gammab _ { k } , \beta_k \}_{k=1}^K$, one can transform the problem as follows. First, we describe $K$ half-spaces with the hyperplanes $\omegab _ { k } ^ { T } \boldsymbol { x } \geq \tau _ { k }$, where
\begin{equation}
\omegab _ { k } = \frac { \gammab _ { k } ^ { T } \Sigmab ^ { 1 / 2 } } { \sqrt { \gammab _ { k } ^ { T } \Sigmab \gammab _ { k } } } , \quad \text { and } \quad \tau _ { k } = \frac { \beta _ { k } - \gammab _ { k } ^ { T } \mub } { \sqrt { \gammab _ { k } ^ { T } \Sigmab \gammab _ { k } } },
\end{equation}
for $k=1,...,K$. Then, a Gaussian $\mcN(\textbf{0},\textbf{I})$ is integrated over the union of those half-spaces as described in previous section.
\newpage
\bibliographystyle{IEEEtran}


\end{document}